# Возбуждение лазеров на парах меди прямым разрядом накопительного конденсатора через быстродействующие фототиристоры

А. С. Кюрегян

ПАО "НПО "ЭНЕРГОМОДУЛЬ", 109052, Москва, Россия
E-mail: semlab@yandex.ru

Исследована возможность применения оптопары «импульсный волоконный лазер – фототиристор» в качестве коммутатора в схемах возбуждения лазеров на парах меди (ЛПМ). Показано, что такой коммутатор обладает наносекундным быстродействием, способен формировать монополярные и знакопеременные импульсы тока через ЛПМ с мощностью до 10 МВт и частотой повторения десятки килогерц при электрическом КПД схемы возбуждения более 95%. Предложена простая, но весьма точная модель фототиристора, которую можно использовать в полномасштабных программах моделирования ЛПМ.

## Введение

Для эффективной работы лазеров на парах меди (ЛПМ) и других металлов необходимо, чтобы схема возбуждения обеспечивала возможно более короткие (не более 10 нс) фронты импульсов напряжения $U_R(t)$ на активной составляющей $R_l(t)$ импеданса газоразрядной трубки [1-3]. Простейшая схема возбуждения изображена на Рис. 1 [1]. В качестве коммутатора в большинстве работ использовались импульсные водородные тиратроны [4], которые обладают рядом недостатков [1,5]. Во-первых, типичная предельная скорость нарастания тока тиратронов $dJ/dt \approx 4$ A/ns $<< U_{C0}/L \approx 50$ A/ns ($U_{C0}$ - начальное напряжение на накопительной емкости $C$, $L$ - индуктивность разрядного контура), а характерное время спада напряжения на тиратроне при включении порядка нескольких десятков наносекунд. Поэтому длительность фронта импульса возбуждения определяется не только параметрами схемы, но и в значительной степени свойствами тиратрона. Во-вторых, потери в тиратронах достигают $(0.4 - 0.6)CU_{C0}^2/2$ [5], что существенно снижает практический КПД всего устройства. В-третьих, ресурс их работы (порядка 1000 часов) недостаточен для ряда практических применений. Поэтому на протяжении уже десятков лет исследуются возможности замены тиратронов на более эффективные коммутаторы, однако, без особого успеха.

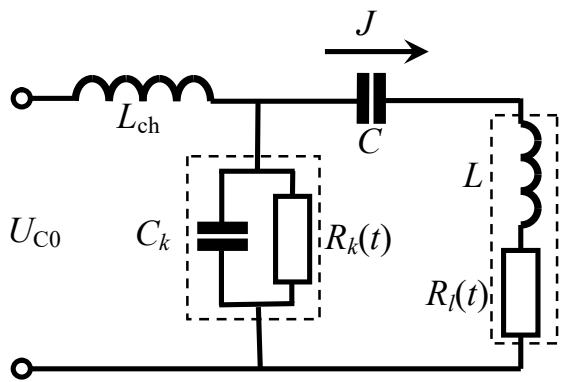

Рис. 1. Эквивалентная схема возбуждения лазера с прямым разрядом накопительной емкости $C$ через коммутатор с емкостью $C_k$ и переменным сопротивлением $R_k(t)$. $L_{ch}$ - индуктивность зарядного дросселя, $L$ - полная индуктивность разрядного контура, $R_l(t)$ - активная составляющая импеданса газоразрядной трубки лазера.

Авторы недавней работы [6], анализируя причины застоя в развитии лазеров на парах металлов за последние 20 лет, пришли к выводу о том, «что выходная мощность и эффективность генерации современных промышленных лазеров ограничиваются несовершенством их систем питания». Они возлагают надежды на использование нового типа разрядников – кивотронов [7,8], которые позволяют резко сократить длительность фронта импульсов возбуждения ЛПМ и тем самым обеспечить «повышение эффективности генерации более чем на порядок при значительном росте выходной мощности». Однако таким образом не решаются другие важные проблемы - повышение КПД и долговечности схемы возбуждения, так как, во-первых, использование кивотрона вносит весьма значительные дополнительные потери и, во-вторых, кивотрон является лишь обострителем фронта импульса, формируемого первичным коммутатором, в качестве которого приходится использовать все тот же тиратрон [6-8].

Между тем еще 37 лет назад авторы работы [9] предложили использовать для возбуждения ЛПМ коммутатор, состоящий из нескольких последовательно соединенных кремниевых фототиристоров специальной конструкции [10], управляемых импульсами излучения неодимового лазера. Однако ни экспериментальных, ни теоретических обоснований эффективности этого технического решения в работе [9] не было представлено и развитие подобных методов возбуждения лазеров на парах металлов приостановилось до сих пор по двум причинам. Во-первых, надежность, длительность импульсов, и КПД стандартных для того времени неодимовых лазеров (типа ЛТИПЧ) были далеко недостаточны для обеспечения необходимых эксплуатационных характеристик фототиристорных коммутаторов. Во-вторых, до недавнего времени фактически отсутствовала теория переключения высоковольтных структур с p-n-переходами из блокирующего в проводящее состояние при воздействии коротких импульсов ионизирующего облучения. Это не позволяло оптимизировать конструкцию фототиристора. Недавно такая теория была построена в работах автора [11-14], результаты которых указывают на возможность переключения высоковольтных кремниевых фототиристоров специальной конструкции в проводящее состояние за время 0,1-5 нс под действием импульсов света с длиной волны $\lambda \approx 1060$ нм, длительностью $t_{ph} = 0.1 - 5$ нс, энергией $W_{ph} = 0.1 - 1$ мДж и частотой повторения $f$ до 80 кГц. Такие параметры управляющих импульсов легко получить с помощью современных коммерческих волоконных лазеров с практическим КПД 10-30% [15].

Столь радикальное улучшение ситуации стимулировало нас заново проанализировать перспективность замены тиратронов на оптопару «волоконный лазер - фототиристор» в схемах возбуждения ЛПМ и других лазеров на самоограниченных переходах. Результаты этого анализа, изложенные далее, показывают, что такая оптопара является почти идеальным коммутатором, обладающим субнаносекундным быстродействием, близким к 100% КПД и высокой долговечностью, свойственной всем твердотельным приборам.

**Модели ЛПМ и коммутатора.**

В настоящей работе мы ограничились рассмотрением чисто электротехнической эффективности работы фототиристоров в качестве коммутаторов схемы возбуждения ЛПМ. Для решения этой задачи нет необходимости использовать полномасштабные модели газоразрядных процессов в ЛПМ, подробно описанные, например, в монографии [1]. Достаточно лишь знать правдоподобные значения индуктивности $L$ и активной составляющей импеданса газоразрядной трубки $R_l(t) = \Lambda / q \mu n \pi r_{pl}^2$, которая зависит от времени $t$ вследствие изменения подвижности $\mu$ и концентрации $n$ свободных электронов в газоразрядной плазме в процессе возбуждения ($\Lambda$ - межэлектродное расстояние газоразрядной трубки, $r_{pl}$ - эффективный радиус плазменного шнура). Мы использовали для этого предельно упрощенную нульмерную модель ЛПМ полагая, что распределение электронов по энергии в первом приближении квазистационарно по отношению к изменению напряженности поля $E = U_R / \Lambda$, где $U_R = R_l J$, $J$ - ток контура. Если рекомбинация электронов во время импульса возбуждения пренебрежимо мала, то в этом случае

$$R_l = \Lambda M(E) / q\mu_0 n S_{pl}, \qquad (1)$$

$$\frac{dn}{dt} = n(N_{Cu} - n) K_{i0} A(E), \quad n(0) = n_0 \qquad (2)$$

где $n_0$ и $\mu_0$ - концентрация и подвижность свободных электронов в момент начала импульса возбуждения, $N_{Cu}$ концентрация атомов меди, $K_{i0} A(E)$ - константа их ионизации, а $M(E)$ и $A(E)$ - мгновенные функции напряженности поля. Для них мы использовали аппроксимации

$$M(E) = \left[1 + (E/E_M)^4\right]^{1/16} \text{ и } A(E) = \exp(-E_A/E),$$

которые неплохо описывают результаты расчетов [16] при значениях $\mu_0 N_{Ne} = 6.6 \cdot 10^{22}\,\text{В}^{-1}\text{с}^{-1}\text{см}^{-1}$, $E_M/N_{Ne} = 2.12 \cdot 10^{-12}\,\text{В}\cdot\text{см}^2$, $K_{i0} = 3 \cdot 10^{-8}\,\text{см}^3\text{с}^{-1}$ и $E_A/N_{Ne} = 1.36 \cdot 10^{-11}\,\text{В}\cdot\text{см}^2$.

Коммутатор представляет собой сборку из $m$ последовательно соединенных кремниевых $n^+$-$p$-$i$-$n$-$p^+$-структур фототиристоров площадью $S_k$, помещенных между молибденовыми электродами (см. Рис. 2). В катодном электроде и в металлизации $n^+$-слоя проделано множество (~100) окон с общей площадью $S_0 \sim 0.5 S_k$ для квазиоднородного освещения. Один из возможных вариантов конструкции такого фототиристора описан в работе [17].

В блокирующем состоянии практически все напряжение $U_{k0}$ падает на встроенных обратно смещенных $p$-$i$-$n$-диодах, толщина которых $d$ незначительно меньше толщины всей кремниевой структуры. До начала освещения через фототиристоры протекает ток утечки

$$J_0 = q n_{k0} v(E_{k0}) S_k, \qquad (3)$$

где $q$ - элементарный заряд, $n_{k0}$ - концентрация электронов и дырок в истощенном $i$-слое, поставляемых инжекцией из эмиттеров и термогенерацией в $i$-слое, $E_{k0} = U_{k0}/md$, $v(E)$ - суммарная дрейфовая скорость электронов и дырок, зависящая от напряженности поля $E$ по закону

$$v(E) = v_{sn} \frac{E}{|E| + E_{sn}} + v_{sp} \frac{E}{|E| + E_{sp}}, \qquad (4)$$

$E_{sn,sp} \ll E_{k0}$ - характерные напряженности поля, выше которых дрейфовые скорости приближаются к величинам $v_{sn,sp} \approx 10^7$ см/с. Для определенности будем считать[1], что каждый фототиристор освещается импульсом света с мощностью

$$P(t) = P_M \frac{\sqrt{2e}}{m} \frac{t}{t_{ph}} \exp\left[-\left(\frac{t}{t_{ph}}\right)^2\right], \qquad (5)$$

где $P_M$ - пиковая мощность. Энергия $m$ таких импульсов $W_{ph} = \sqrt{e/2}\, P_M t_{ph}$. После начала освещения усредненная по толщине структуры концентрация $\bar{n}_k$ в освещенных областях увеличивается со временем по закону

$$\bar{n}_k(t) = n_{k0} + n_{k1}\left\{1 - \exp\left[-(t/t_{ph})^2\right]\right\}, \qquad n_{k1} = W_{ph} \frac{1 - R_{ph}}{\hbar\omega S_0} \frac{1 - e^{-\kappa d}}{md}, \qquad (6)$$

где $\hbar\omega$ - энергия кванта света, $R_{ph}$ - коэффициент отражения от поверхности окна, $\kappa$ - коэффициент поглощения свет в кремнии. Можно показать, что при этом падение напряжения $U_k$ на коммутаторе связано с протекающем через нагрузку током $J$ соотношением [11]

$$C_k \frac{dU_k}{dt} = J - S_0 \bar{j}, \qquad (7)$$

---
[1] Окончательные результаты очень слабо зависят от формы импульса излучения.

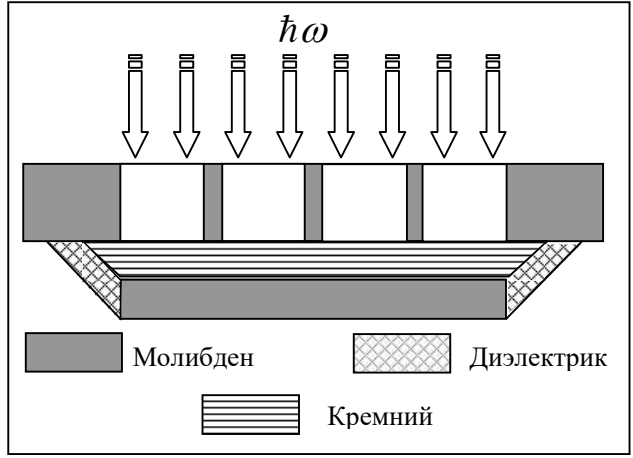

Рис. 2. Схематичное изображение поперечного сечения фототиристора.

где $C_k = \varepsilon S_k / md$ - емкость коммутатора в блокирующем состоянии, $\varepsilon$ - диэлектрическая проницаемость кремния, $\bar{j}$ - усредненная по толщине $i$-слоя плотность тока электронов и дырок в освещенной области. Строго говоря, для вычисления функции $\bar{j}(t)$ нужно решить уравнение (7) вместе с уравнением Пуассона и уравнениями непрерывности для концентраций электронов и дырок, как это сделано в [11]. Однако в настоящей работе мы используем упрощенную формулу

$$\bar{j} = q\bar{n}_k v(E_k), \qquad (8)$$

которая приближенно верна при $t \ll d/v$ - времени пролета электронов и дырок через $i$-слой, если напряженность поля в $i$-слое слабо зависит от координаты (то есть $E_k \approx U_k/md$) и $\kappa d < 1$. Эта упрощенная модель оправдана (и точное численное моделирование – см. далее – подтверждает ее применимость) не только потому, что в качественном отношении правильно описывает переключение фототиристоров в проводящее состояние, но и вследствие слабого влияния параметров «почти идеального коммутатора» на процесс возбуждения ЛПМ. Необходимо подчеркнуть, что сказанное относится только к приборам специальной конструкции [10,17] при надлежащем выборе параметров управляющих импульсов света, но совершенно неприменимо к обычным высоковольтным фототиристорам [18].

Используя законы Кирхгофа для контура и формулы (1)-(8) нетрудно получить следующую систему уравнений и соответствующих им начальных условий:

$$\frac{dJ}{dt} = \frac{Q}{\tau^2} - \frac{U_R + U_k}{\rho\tau}, \qquad J(0) = J_0, \qquad (9)$$

$$\frac{dQ}{dt} = -J, \qquad Q(0) = CU_{C0}, \qquad (10)$$

$$\frac{dn}{dt} = n(N_{Cu} - n)K_{i0}A(U_R/\Lambda), \qquad \eta(0) = 0 \qquad (11)$$

$$\frac{dU_R}{dt} = R_0 M \frac{\dfrac{dJ}{dt} - \dfrac{J}{n}\dfrac{dn}{dt}}{\dfrac{n}{n_0} - \dfrac{JR_0}{\Lambda}\dfrac{dM}{dE}}, \qquad U_R(0) = J_0 R_0, \qquad (12)$$

$$\frac{dU_k}{dt} = C_k^{-1}\left[J - q\bar{n}_k v\left(\frac{U_k}{md}\right)S_0\right], \qquad U_k(0) = U_{k0}, \qquad (13)$$

где $\tau = \sqrt{LC}$, $\rho = \sqrt{L/C}$, $R_0 = \Lambda/q\mu_0 n_0 S_{pl}$, $Q$ - заряд накопительной емкости, $U_{C0} = (U_{k0} + U_{R0})$ - начальное напряжение на накопительной емкости.

**Результаты расчетов и моделирования**

В этом разделе в качестве примера изложены результаты решения системы приближенных уравнений (9)-(13) методом Рунге-Кутта и точного численного моделирования процессов в фототиристоре с помощью программы «Исследование» [19]. Использовались значения параметров ЛПМ и контура, взятые из работы [3]: $\Lambda = 48$ см, $r_{pl} = 1$ см, $N_{Ne} = 2 \cdot 10^{18}$ см$^{-3}$, $N_{Cu} = 1.5 \cdot 10^{-3} N_{Ne}$, $n_0 = (2-20) \cdot 10^{13}$ см$^{-3}$, индуктивность газоразрядной трубки с коаксиальным обратным токопроводом $L = 160$ нГн, накопительная емкость $C = 1.5$ нФ заряжена до начального напряжения $U_{C0} = 15$ кВ. Предполагалось, что коммутатор состоит из $m = 3$ последовательно соединенных кремниевых фототиристоров со следующими параметрами каждого из них: $S_k = 1$ см$^2$, $S_0 = 0.5 S_k$, $d = 600$ мкм, напряжение пробоя $U_b = 6.34$ кВ, ток утечки $J_0 = 1.7$ мА при $U_{k0} = 5$ кВ, время восстановления блокирующей способности 1-10 мкс. Тиристоры переключаются в проводящее состояние под действием импульсов света с длиной волны 1,064 мкм (коэффициент поглощения в кремнии $\kappa \approx 32$ см$^{-1}$ при рабочей температуре кристалла 100 $^0$С).

Результаты расчетов приведены на Рис. 3-5. Зависимости $R(t)$, примеры которых изображены на Рис. 3, использовались в программе «Исследование» [19] для точного моделирования процессов в тиристорных $n^{++} - p^+ - p - n - n^+ - p^{++} -$структурах, основные параметры которых приведены выше, а реалистичное распределение легирующих примесей по толщине кристаллов описано в [12]. Полученные таким образом «приближенные» и «точные» зависимости $U_k(t)$ и $U_R(t)$ хорошо согласуются между собой (см. Рис.4). Заметное относительное расхождение наблюдается только при очень малых (порядка 1 В) значениях $U_k(t)$. Причина этого, очевидно, состоит в том, что упрощенная модель тиристоров не учитывает падения напряжения (около 0,7 В) на прямо смещенных эмиттерных переходах. Однако такое расхождение практически не влияет на результаты расчетов энергии потерь в тиристорах $W_k$ (см. Рис. 5). Как видно, $W_k$ начинает резко увеличиваться с ростом длительности управляющего импульса света при $t_{ph} > 10$ нс. Поэтому для управления фототиристорами следует использовать импульсы света с $t_{ph} \leq 5$ нс, обеспечивающие переключение в проводящее состояние за время порядка 1 нс или меньше.

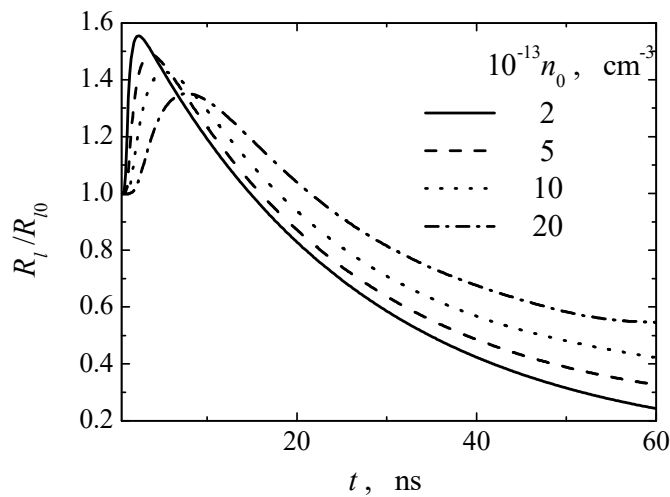

Рис. 3. Зависимости активной составляющей импеданса лазера $R_l$ от времени, полученные путем решения системы приближенных уравнений (9)-(13) при $W_{ph} = 3 \times 108$ мкДж, $t_{ph} = 5$ нс и различных предимпульсных концентрациях электронов $n_0$.

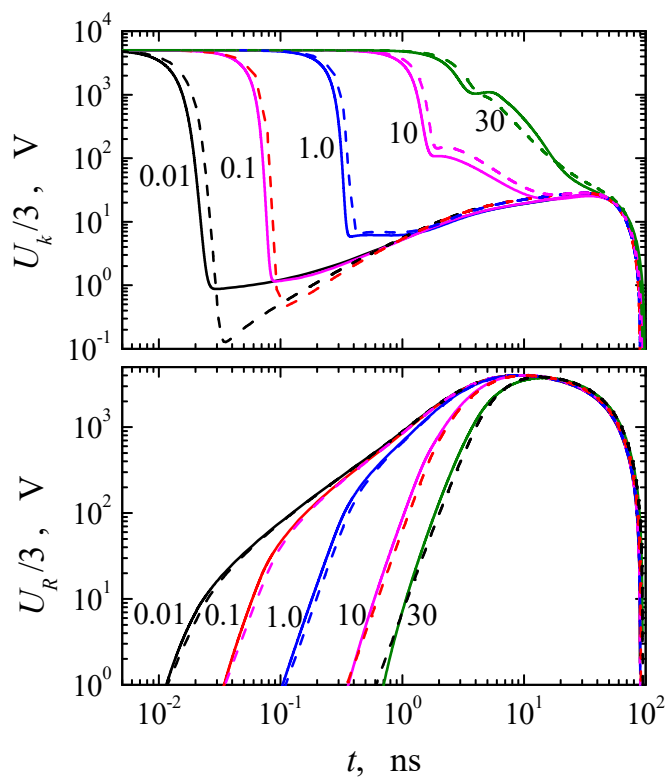

Рис. 4. Зависимости напряжений на коммутаторе $U_k$ и на активной составляющей импеданса лазера $U_R$ от времени, полученные путем точного численного моделирования (сплошные линии) и решения системы уравнений (9)-(13) (штриховые линии) при $W_{ph} = 3 \times 122$ мкДж и $n_0 = 10^{14}$ см$^{-3}$. Цифры у кривых означают длительность управляющего импульса света $t_{ph}$ в наносекундах.

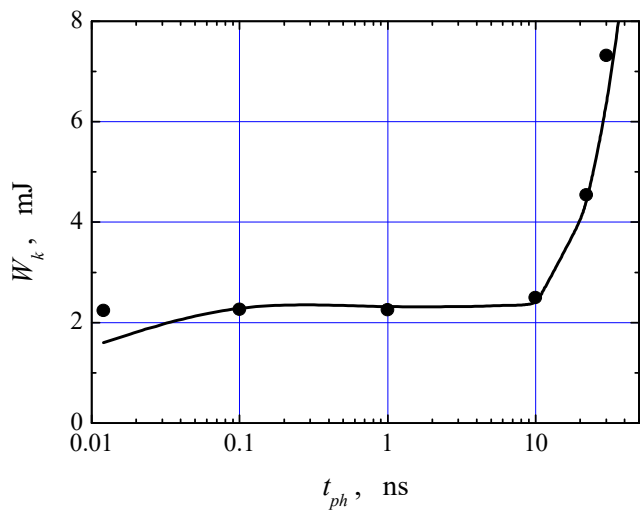

Рис. 5. Зависимости энергии коммутационных потерь в трех фототиристорах $W_k$ от длительности управляющего импульса света $t_{ph}$, полученные путем точного численного моделирования (символы) и путем решения системы приближенных уравнений (9)-(13) (линия) при $W_{ph} = 3 \times 200$ мкДж и $n_0 = 10^{14}$ см$^{-3}$.

Энергия потерь в тиристорах $W_k$ уменьшается с ростом $W_{ph}$, но одновременно увеличиваются потери в управляющем лазере. В результате получается немонотонная зависимость суммарной энергии потерь в оптопаре $W_{sum}$ от $W_{ph}$, которая имеет минимум (см. Рис. 6). Таким образом, существует оптимальное значение $W_{ph}$, которое в нашем конкретном случае равно 0,33 мДж при $t_{ph} = 5$ нс и КПД управляющего лазера 10%, когда $W_{sum} \approx 2W_k \approx 6.6$ мДж. Так как запасенная в накопительной емкости $C$ энергия $CU_0^2/2 = 169$ мДж, то без учета потерь в цепи заряда получается КПД схемы возбуждения более 96%. Импульсный разогрев тиристоров за один цикл возбуждения равен примерно 0,01 К, так что фактически они работают в режиме постоянного рассеивания средней мощности $P_k = W_k f$. При $f = 20$ кГц каждый из трех тиристоров будет рассеивать среднюю мощность $P_k/m \approx 22$ Вт, поэтому для поддержания рабочей температуры кристаллов $100^\circ$C достаточно использовать простую систему охлаждения с тепловым сопротивлением кристалл-среда 4 К/Вт.

Форма и амплитуда импульсов тока $J(t)$ контура и напряжения $U_R(t)$ на активной составляющей импеданса ЛПМ остаются практически постоянными при изменении длительности и энергии управляющих импульсов света в разумных пределах, но существенно зависят от предимпульсной концентрации электронов $n_0$. Вследствие этого изменяется зависимость мощности $P_R = U_R J$ от времени и энергия $W_R$, поглощаемая ЛПМ на начальном этапе возбуждения, как это изображено на Рис. 7,8. Кроме этого наблюдается обратное влияние ЛПМ на коммутатор: с ростом $n_0$ увеличиваются потери энергии в фототиристорах (см. Рис. 8).

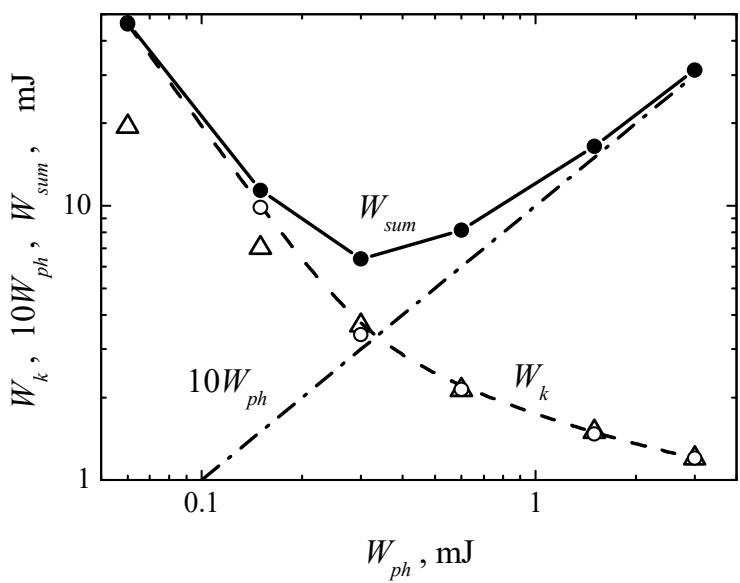

Рис. 6. Зависимости энергии, рассеиваемой тремя тиристорами коммутатора $W_k$, полученые путем точного численного моделирования (светлые кружки + штриховая линия) и путем решения системы приближенных уравнений (9)-(13) (светлые треугольники)), источником питания управляющего лазера с КПД 10% ($10W_{ph}$, штрих-пунктир) и суммарной энергии потерь оптопары $W_{sum}$ за один импульс возбуждения (темные кружки + сплошная линия) от энергии $W_{ph}$ импульса излучения управляющего лазера при $t_{ph} = 5$ нс и $n_0 = 10^{14}$ см$^{-3}$.

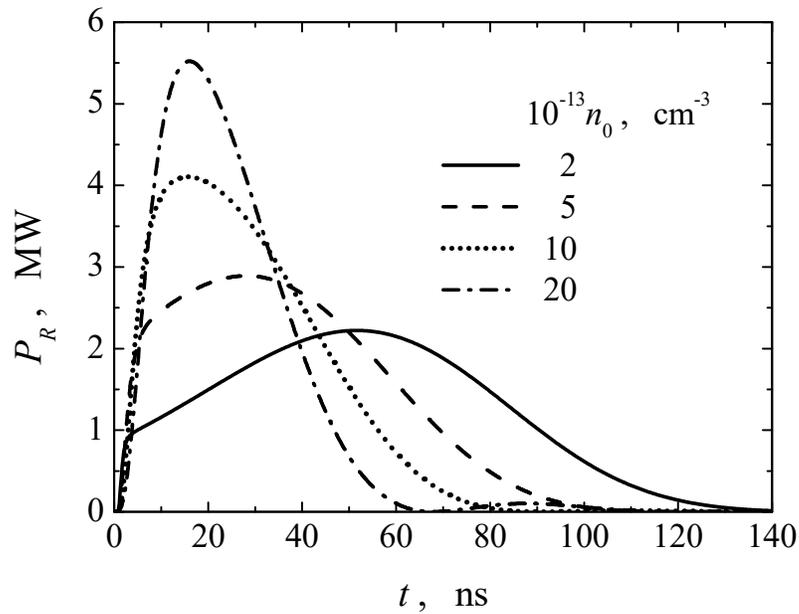

Рис. 7. Зависимости мощности $P_R$, рассеиваемой активной составляющей импеданса лазера $R_l$ от времени, полученные путем решения системы приближенных уравнений (9)-(13) при $W_{ph} = 3\times 108$ мкДж, $t_{ph} = 5$ нс и различных предимпульсных концентрациях электронов $n_0$.

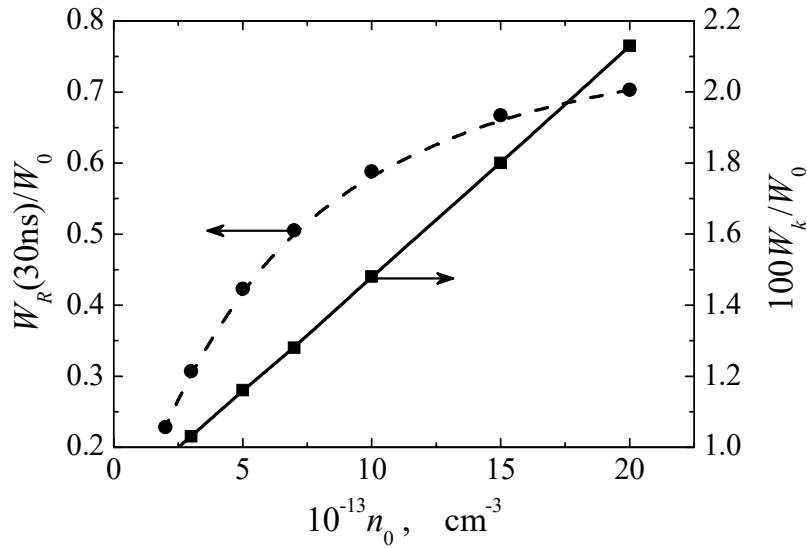

Рис. 8. Зависимости нормированных значений энергии $W_R$, рассеиваемой активной составляющей импеданса лазера $R_l$ за 30 нс после начала импульса возбуждения, и энергии $W_k$, рассеиваемой коммутатором, от предимпульсной концентрациии электронов $n_0$, полученные путем решения системы приближенных уравнений (9)-(13) при $W_{ph} = 3\times 108$ мкДж и $t_{ph} = 5$ нс.

В рассмотренном нами примере контур возбуждения оказался почти точно согласованным с нагрузкой: амплитуда отраженного импульса была примерно в 10 раз меньше, чем основного. Однако на практика такого согласования удается достичь далеко не всегда и возникновение мощного отраженного импульса приводит к дополнительному ужесточению режима работы тиратронов [1,5]. В связи с эти следует отметить еще одно достоинство описанного коммутатора: фототиристоры способны пропускать большие обратные токи в течении сотен наносекунд безо всякого для себя вреда. Разумеется, при этом увеличиваются потери, но незначительное увеличение энергию управляющего импульса свет позволяет полностью скомпенсировать этот эффект. Поэтому фототиристоры можно использовать почти столь же эффективно даже для возбуждения коаксиальных ЛПМ импульсно-периодическим индукционным разрядом [20], который невозможно реализовать при использовании тиратронов.

В заключение отметим три проблемы, связанные с возможностью применения описанного коммутатора. Первая из них – сложность создания малоиндуктивной сборки последовательно соединенных фототиристоров [9] и отсутствие промышленной технологии изготовления кремниевых структур с напряжением пробоя $U_b$ более 10 кв [18]. В принципе ее можно решить, используя карбид-кремниевые фототиристоры, так как уже созданы опытные образцы биполярных ключей на основе 4H-SiC с $U_b = 27$ кВ [21]. Однако промышленная технологии подобных приборов также еще не освоена, а управляющие лазеры с подходящей длиной волны излучения $\lambda = 375$ нм труднодоступны. Поэтому в настоящее время описанный коммутатор можно применять для возбуждения либо не очень мощных ЛПМ с короткой газоразрядной трубкой, либо многосекционных ЛПМ [22,23] и ЛПМ с поперечным разрядом [24,25]. Вторая проблема – высокая стоимость волоконных лазеров, которая может решиться в ходе дальнейшего прогресса технологии импульсных полупроводниковых лазеров. Третья проблема состоит в том, что изложенные выше результаты, иллюстрирующие достоинства оптопары «волоконный лазер - фототиристор», не позволяют ответить на вопрос о том, насколько таким способом можно улучшить выходные оптические характеристики ЛПМ - среднюю мощность излучения и полный практический КПД. Для этого необходимо решить задачу, выходящую за рамки настоящей работы. Именно надо провести оптимизацию режима работы ЛМП с использованием одной из полномасштабных моделей [1], заменив в них феноменологическое описание тиратрона (см., например, раздел 8.4 в [1]) на весьма точное и физически обоснованное уравнение (13).

**Литература**

1. *Лазеры на самоограниченных переходах атомов металлов*-2. Т. 1, Под редакцией В.М. Батенина (М.: ФИЗМАТЛИТ, 2009, 544 с.)
2. Григорьянц А. Г., Казарян М. А., Лябин Н. А. *Лазеры на парах меди: конструкция, характеристики и применения* (М.: ФИЗМАТЛИТ, 2005, 312 с.)
3. Бохан П.А., Гугин П.П., Закревский Дм.Э., Лаврухин М.А., Казарян М.А., Лябин Н.А. *Квантовая электроника,* **43**, 715 (2013).
4. Фогельсон Т.Б., Бреусова Л.Н., Вагин Л.Н. *Импульсные водородные тиратроны* (М.: Сов. радио, 1974, 212 с.)
5. Исаев А.А., Леммерман Г.Ю. Труды ФИАН, **181**, 64 (1987).
6. Бохан П.А., Закревский Дм.Э., Лаврухин М.А., Лябин Н. А., Чурсин А. Д. *Квантовая электроника,* **46**, 100 (2016).
7. Бохан П.А., Гугин П.П., Закревский Дм.Э., Лаврухин М.А. *Письма в ЖТФ*, **38** (8), 63 (2012).
8. Бохан П.А., Гугин П.П., Закревский Дм.Э., Лаврухин М.А. *Письма в ЖТФ*, **39** (17), 44 (2013).
9. Александров В. М., Бужинский О. И., Грехов И. В. и др., *Квантовая электроника,* **8**, 191 (1981).
10. Воле В. М., Воронков В. М., Грехов И. В. и др., *ЖТФ*, **51**, 373 (1981).
11. Кюрегян А.С., *ФТП*, **48**, 1686 (2014).
12. Кюрегян А.С., *ФТП*, **51**, 1257 (2017).
13. Кюрегян А.С., *ФТП*, **51**, 1263 (2017).
14. Кюрегян А.С., http://arxiv.org/abs/1806.06922
15. http://www.ipgphotonics.com/ru/products/lasers

16. Мнацаканян А. Х., Найдис Г. В., Штернов Н. П. *Квантовая электроника*, **5**, 597 (1978).
17. Glidden S. C., Sanders H D. US Patent № 8,461,620 B2 (2013).
18. http://www.elvpr.ru/poluprovodnikprib/tiristory/flyers/TL_2017.pdf
19. Mnatsakanov T. T., Rostovtsev I. L., Philatov N. I. *Solid-State Electronics*, **30**, 579 (1987).
20. Батенин В.М., Казарян М.А., Карпухин В.Т., Лябин Н.А., Маликов М.М., Сачков В.И. *Оптика атмосферы и океана*, **29**, 112 (2016).
21. Brunt E., Cheng L, O'Loughlin M.J et al, *Materials Science Forum,* **821-823,** 847 (2015).
22. *Pack J.L., Liu C.S., Feldman D.W., Weaver L.A. Rev. Sci. Instrum*., **48**, 1047 (1977).
23. Кирилов А. Е., Кухарев В. Н., Солдатов А. Н., *Квантовая электроника,* **6**, 473 (1979).
24. Piper J. A., *IEEE J. Quantum Electron.*, **QE-14**, 405 (1978).
25. Соколов А. В., Свиридов А. В., *Квантовая электроника*, **8,** 1686 (1981).